\documentclass[aip, amsmath,amssymb, preprint]{revtex4-1}

\usepackage{graphicx}
\usepackage[utf8]{inputenc}
\usepackage[T1]{fontenc}
\usepackage{mathptmx}
\usepackage{color}
\usepackage[normalem]{ulem}
\usepackage{hyperref}
\hypersetup{
    colorlinks=true,
    linkcolor=blue,
    citecolor=magenta,
    urlcolor=cyan,
}
\usepackage{microtype}		

\begin{document}

\title{Weak antilocalization and localization in Eu$_2$Ir$_2$O$_7$ (111) thin films by reactive solid phase epitaxy}
\author{Xiaofeng Wu}
	\affiliation{Beijing National Laboratory for Condensed Matter Physics and Institute of Physics, Chinese Academy of Sciences, Beijing 100190, China.}
	\affiliation{School of Physical Sciences, University of Chinese Academy of Sciences, Beijing 100049, China.}
\author{Zhen Wang}
	\affiliation{Beijing National Laboratory for Condensed Matter Physics and Institute of Physics, Chinese Academy of Sciences, Beijing 100190, China.}
	\affiliation{School of Physical Sciences, University of Chinese Academy of Sciences, Beijing 100049, China.}
	\affiliation{Institute of High Energy Physics, Chinese Academy of Sciences, Beijing 100049, China.}
\author{Zhaoqing Ding}
	\affiliation{Beijing National Laboratory for Condensed Matter Physics and Institute of Physics, Chinese Academy of Sciences, Beijing 100190, China.}
	\affiliation{School of Physical Sciences, University of Chinese Academy of Sciences, Beijing 100049, China.}
\author{Zeguo Lin}
	\affiliation{Beijing National Laboratory for Condensed Matter Physics and Institute of Physics, Chinese Academy of Sciences, Beijing 100190, China.}
	\affiliation{School of Physical Sciences, University of Chinese Academy of Sciences, Beijing 100049, China.}
\author{Mingyu Yang}
	\affiliation{Beijing National Laboratory for Condensed Matter Physics and Institute of Physics, Chinese Academy of Sciences, Beijing 100190, China.}
	\affiliation{School of Physical Sciences, University of Chinese Academy of Sciences, Beijing 100049, China.}
\author{Minghui Gu}
	\affiliation{Beijing National Laboratory for Condensed Matter Physics and Institute of Physics, Chinese Academy of Sciences, Beijing 100190, China.}
	\affiliation{School of Physical Sciences, University of Chinese Academy of Sciences, Beijing 100049, China.}
\author{Meng Meng}
	\affiliation{Beijing National Laboratory for Condensed Matter Physics and Institute of Physics, Chinese Academy of Sciences, Beijing 100190, China.}	
\author{Fang Yang}
	\affiliation{Beijing National Laboratory for Condensed Matter Physics and Institute of Physics, Chinese Academy of Sciences, Beijing 100190, China.}	
\author{Xiaoran Liu}
	\email{xiaoran.liu@iphy.ac.cn}
	\affiliation{Beijing National Laboratory for Condensed Matter Physics and Institute of Physics, Chinese Academy of Sciences, Beijing 100190, China.}
\author{Jiandong Guo}
	\email{jdguo@iphy.ac.cn}
	\affiliation{Beijing National Laboratory for Condensed Matter Physics and Institute of Physics, Chinese Academy of Sciences, Beijing 100190, China.}
	\affiliation{School of Physical Sciences, University of Chinese Academy of Sciences, Beijing 100049, China.}

\date{\today}

\begin{abstract}

Thin films of the pyrochlore iridates along the [111] direction have drawn significant attention to investigate exotic correlated topological phenomena. Here, we report the fabrication of Eu$_2$Ir$_2$O$_7$ thin films via reactive solid phase epitaxy using the pulsed laser deposition technique. We mainly focus on the transport properties of the films below the magnetic phase transition at 105 K. Analyses on the temperature and the field dependences of resistivity unveil the presence of weak antilocalization, a characteristic signature of the Weyl semimetallic state that has been ``buried'' by magnetism. Moreover, it is noteworthy that the contribution from many-body interactions in Eu$_2$Ir$_2$O$_7$ thin films is enhanced at lower temperatures and competes with the weak antilocalization effect, and eventually drives the crossover to weak localization at 2 K.\\ 

Authors to whom correspondence should be addressed: [Xiaoran Liu, xiaoran.liu@iphy.ac.cn; Jiandong Guo, jdguo@iphy.ac.cn]

\end{abstract}

\maketitle


\begin{figure*}[htp]
\begin{center}
\includegraphics[width=.80\textwidth]{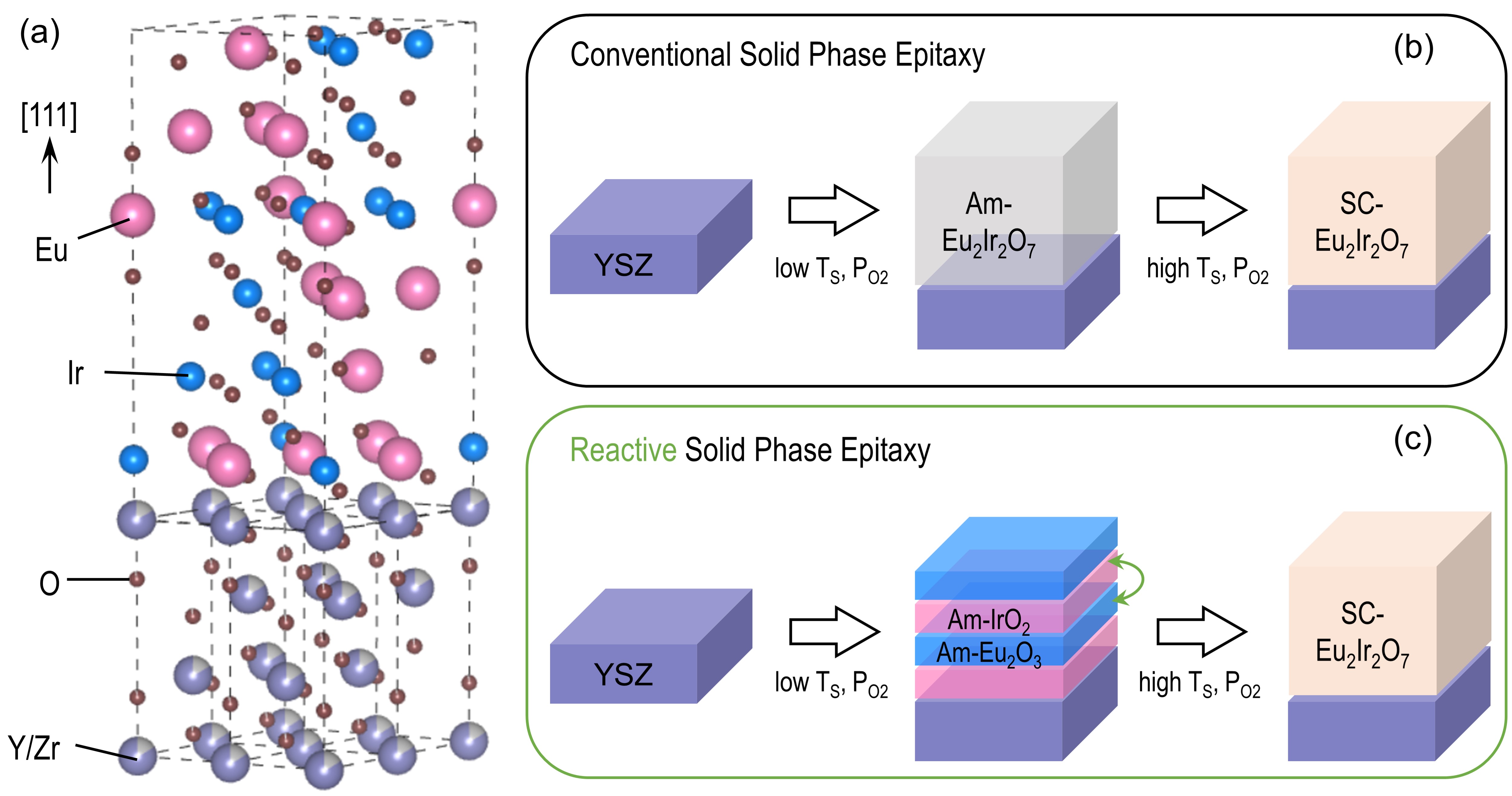}
\caption{\label{Fig1}(a) Illustration about the crystal structure of epitaxial Eu$_2$Ir$_2$O$_7$ films on (111) YSZ substrate. (b)-(c) Schematic descriptions on the deposition process of (b) conventional solid phase epitaxy and (c) reactive solid phase epitaxy for the synthesis of epitaxial Eu$_2$Ir$_2$O$_7$ films. T$_S$ and P$_{\textrm{O}_2}$ denote for the temperature of substrate and the pressure of O$_2$ atmosphere; `Am' and `SC' stand for amorphous and single-crystalline, respectively.}
\end{center}
\end{figure*}

Complex oxides with 5$d$ transition metal elements are intriguing systems, where the interplay of strong spin-orbit coupling (SOC) and intermediate electron-electron interactions can lead to numerous exotic quantum states of matter. \cite{witczak2014correlated,schaffer2016recent} 
In this context, the pyrochlore iridates R$_2$Ir$_2$O$_7$ (R = Y or rare-earth element) have been extensively investigated from theory over the past decade, as a promising playground to realize a variety of fascinating correlated and topological phenomena, such as topological insulators\cite{pesin2010mott,yang2010topological}, Weyl metals and semimetals\cite{wan2011topological,wang2017prb}, axion insulators\cite{varnava2018prb}, topological magnons\cite{laurell2017prl}, magnetic multipoles\cite{suzuki2019prb}, quantum spin liquids\cite{chen2016prb}. In particular, Eu$_2$Ir$_2$O$_7$ (EIO) has attracted great interest to investigate the intrinsic properties of Ir$^{4+}$ on the pyrochlore sublattice, as the Eu$^{3+}$ ions are non-magnetic. Earlier results from EIO powders revealed a metal-nonmetal transition below 120 K, concurrently with a para- to antiferro-magnetic (AFM) transition \cite{daiki2001jpsj, taira2001magnetic}. In the AFM phase, the local Ir moments form the noncoplanar ``all-in-all-out'' configuration, corresponding to a peculiar magnetic octupolar order \cite{liang2017np}. This magnetic octupole can break the time-reversal symmetry and trigger the formation of multiple pairs of Weyl nodes at the Fermi energy \cite{wan2011topological}. Recently, indications of the Weyl semimetal (WSM) state in EIO have been uncovered by a growing number of experiments \cite{Liu2021prl, sushkov2015prb, Tafti2012prb}. Nevertheless, the presence of electronic correlations can lead to substantial many-body effects, such that the unique properties generated from band topology might be suppressed or overwhelmed \cite{Hohenadler2013jpcm, Rachel2018rpp,Roy2017prb}. As the result, the characteristic signatures for WSM in such an interacting system still remain largely elusive. 

Moreover, the experimental investigation has been hindered due to great challenges in the synthesis of high-quality EIO single crystals and epitaxial films, where a set of intertwined thermodynamic restrictions lead to a narrow window for the pyrochlore phase \cite{chakhalian2020aplm, kim2019operando, abb2019growth, gong2013thermal, martin2020high, abb2018template, geiger2016activity, fujita2015odd, kim2019operando, hou2017microstructure,guo2020route,el1996pulsed,mahmoud2010electrical,vente1991orthorhombic}. To overcome these issues, people have lately adopted the so-called `solid phase epitaxy' (SPE) method to achieve (111)-oriented single-crystalline EIO epitaxial films \cite{fujita2015odd}. Note, the (111) plane is the natural cleavage plane of pyrochlore, and the crystal structure and epitaxial relationship to the yttria-stabilized zirconia (YSZ) substrate are illustrated in Fig.~\ref{Fig1}(a). Conventionally during the SPE process by pulsed laser deposition techniques as depicted in Fig.~\ref{Fig1}(b), Ir-rich sputtering targets are intentionally used to compensate the iridium loss. EIO amorphous films with flat morphology and proper stoichiometry are first established on substrates, by adjusting the parameters in a relatively milder condition (i.e., low substrate temperature T$_\textrm{S}$ and oxygen partial pressure P$_{\textrm{O}_2}$). Then, the films are annealed under conditions of much higher temperature and oxygen pressure, to crystallizing into the pyrochlore structure. However, it is noteworthy that since the target is premixed with a fixed molar ratio of R:Ir, success of fabrication in this manner has to significantly rely on the precise tuning of T$_\textrm{S}$ and P$_{\textrm{O}_2}$, which usually requires a complicated gas atmosphere (e.g., a mixture of oxygen and argon) \cite{fujita2015odd, fujita2016apl, Liu2021prl}. In this regard, additional route with more controlling knobs is highly demanded.  

In this Letter, we report the fabrication of EIO (111) epitaxial thin films by means of a method developed from the reactive solid phase epitaxy \cite{katayama2019reactive,ohta2002reactive}. The main theme manifests in the alternate ablation of two separate targets, Eu$_2$O$_3$ and IrO$_2$, to establish the amorphous film in pure  O$_2$ atmosphere. The pyrochlore phase is achieved during post-annealing process, via diffusion and reaction of the two precursors through the interfaces. X-ray diffraction (XRD) experiments have demonstrated the proper crystal structure and expected epitaxial relationship without secondary phases. Temperature-dependent resistivity data reveal the onset of a metal to semimetal transition near 105 K. Combined the temperature and field dependences of resistivity unveil the presence of weak antilocalization (WAL) at low temperatures, which is a characteristic signature of WSM that has been shadowed by Ir magnetism. Notably, as temperature further decreases, a crossover from WAL to weak localization (WL) is uncovered below $\sim$5 K plausibly driven by the more dominant contribution from electron-electron interactions, as suggested by theoretical model calculations \cite{lu2015prb}. 

We first introduce more details about the whole process of samples fabrication, as illustrated in Fig.~\ref{Fig1}(c). The films were deposited on 5$\times$5 mm$^2$ YSZ (111) substrates by pulsed laser deposition. Ceramic IrO$_2$ and Eu$_2$O$_3$ targets were sintered by solid state reaction, and ablated inside the chamber using a KrF excimer laser ($\lambda$ = 248 nm). During the first stage of making amorphous films, the laser fluence was selected at $\sim$1.0 J/cm$^2$ with a fixed repetition rate of 1 Hz. To explore the proper growth conditions, T$_S$ was varied from room temperature to 650 $^\circ$C, and P$_{\textrm{O}_2}$ from 10$^{-7}$ to 10$^{-1}$ Torr. The growth rate of each IrO$_2$ and Eu$_2$O$_3$ layer was calibrated individually at each condition in terms of pulses per nanometer, with the optimized number of pulses at 55 (30) for IrO$_2$ (Eu$_2$O$_3$). After cooled to room temperature, the films were then transferred into a muffle furnace and annealed in air for a few hours, with the annealing temperature varied from 800 $^\circ$C to 1200 $^\circ$C. 

\begin{figure}[htp]
\begin{center}
\includegraphics[width=.65\textwidth]{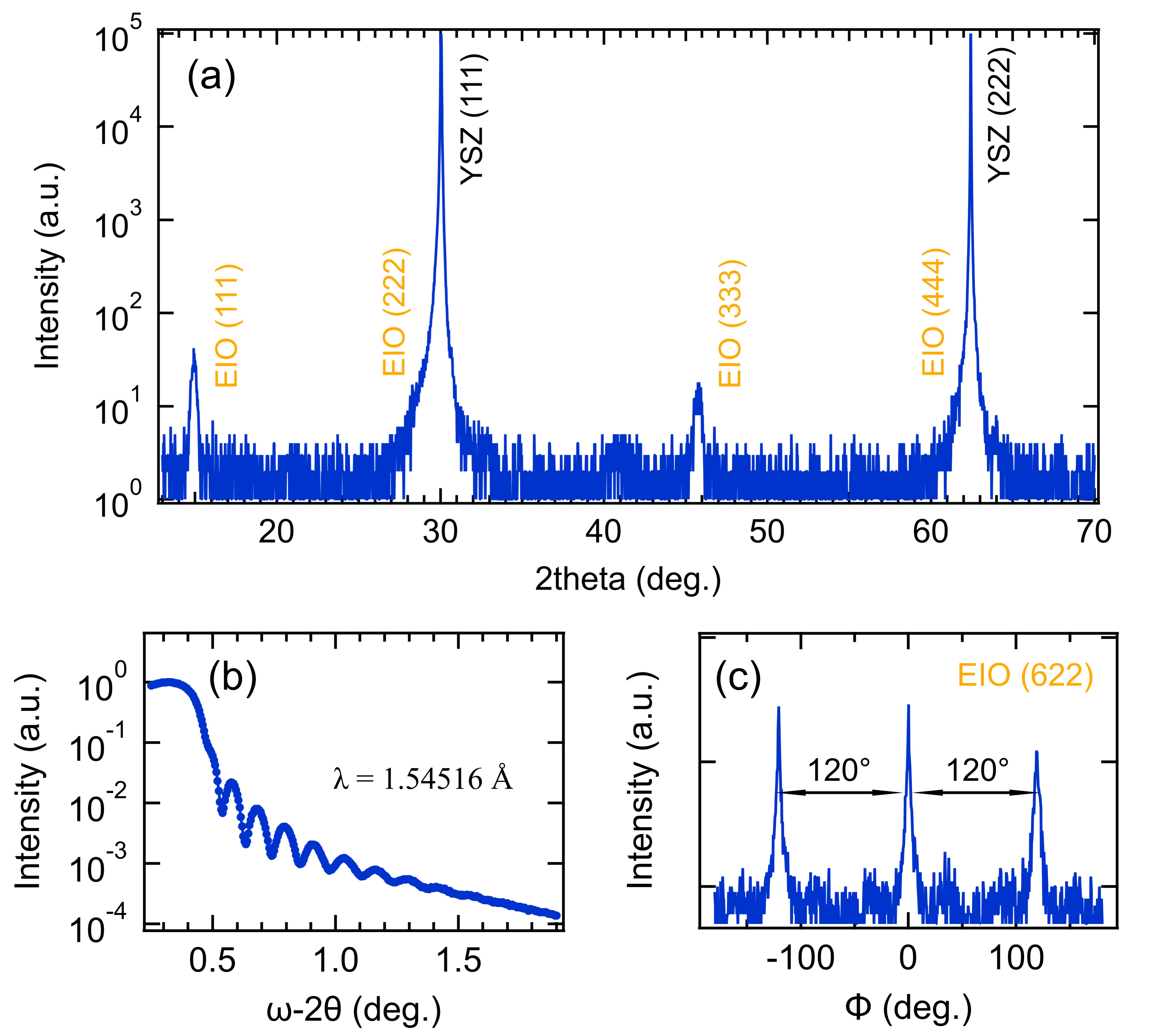}
\caption{\label{Fig2} (a) Wide angle x-ray diffraction symmetric scans of the EIO film on the (111) YSZ  substrate. The Bragg reflections of YSZ (in black) and EIO (in orange) are labeled on the figure. (b) X-ray reflectivity scan of EIO film. (c) Azimuthal $\Phi$-scan measurement around the EIO (622) reflection.}
\end{center}
\end{figure}

We find that single phase EIO epitaxial films can be achieved in a window spanned by T$_S$ (450 - 550 $^\circ$C) and P$_{\textrm{O}_2}$ (20 - 40 mTorr). A representative XRD 2$\theta$-$\omega$ scan of these films is shown in Fig.~\ref{Fig2}(a). The odd-index reflections of the pyrochlore structure (i.e., EIO (111) and (333) as labeled on the figure) are clearly observed, giving rise to an estimation of the out-of-plane spacing $d_{111}$ of $\sim$5.92 \AA, practically the same as EIO bulk. Thickness of the films is around 40 nm, as extracted from the x-ray reflectivity scans (Fig.~\ref{Fig2}(b)), which is in accord with the values estimated based on the total number of pulses. In addition, in the vicinity of YSZ (311) reflection, the EIO (622) reflection is detected, confirming the expected epitaxial relationship as EIO (111)[11$\bar{2}$] // YSZ (111)[11$\bar{2}$]. The azimuthal $\Phi$ scan around the EIO (622) reflection exhibits the  three-fold symmetry (Fig~\ref{Fig2}(c)), indicating our films contain neither mis-oriented domains nor stacking faults within the experimental precision. 


\begin{figure}[htp]
\begin{center}
\includegraphics[width=0.65\textwidth]{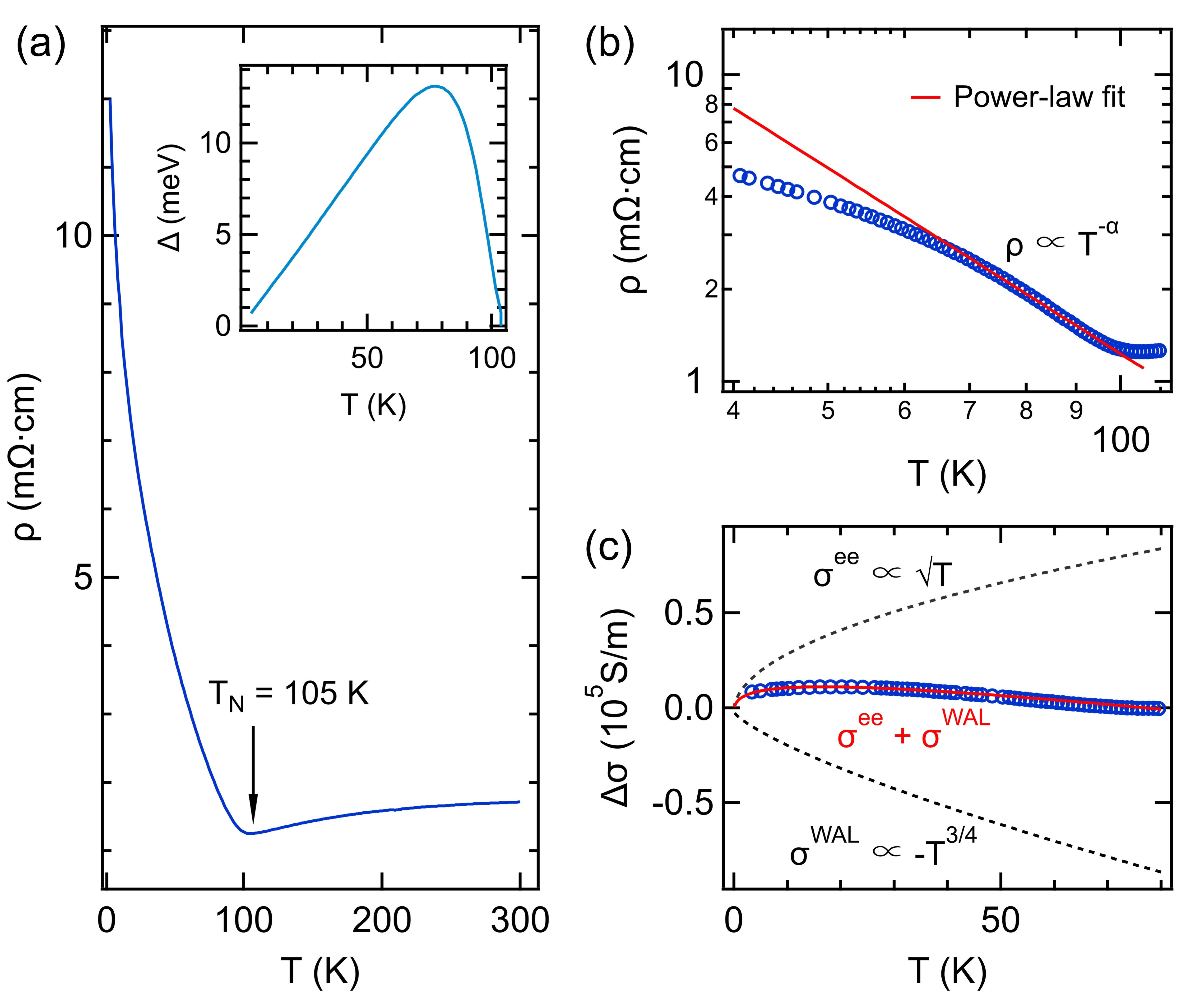}
\caption{\label{Fig3} (a) Temperature-dependent resistivity of (111) EIO thin films. The inset shows the extracted charge gap as a function of temperature by assuming the Arrhenius model. (b) Power-law fit of resistivity data in the range of 100-70 K ($\alpha \approx$ 2.0). (c) Fit (red curve) of the low-temperature conductivity (blue spots) after subtracting the power-law contribution. The net signal contains two terms, $\sigma^\textrm{ee}$ from the electron-electron interaction and $\sigma^\textrm{WAL}$ from the WAL effect.}
\end{center}
\end{figure}

Next, we turn to focus on investigating the electronic properties of EIO films from transport experiments. The temperature dependence of resistivity $\rho(T)$ was measured from 300 K to 3 K in the Van der Pauw geometry. As shown in Fig~\ref{Fig3}(a), the $\rho(T)$ curve exhibits a clear kink around 105 K, signifying the para- to antiferro-magnetic (AFM) phase transition as observed in EIO bulk compound \cite{ishikawa2012continuous}. While a metallic feature is identified above T$_\textrm{N}$, we mainly pay attention to the intriguing behavior below the transition, where $\rho(T)$ increases monotonically as temperature decreases, indicating an insulating or semiconducting feature. However, the $\rho(T)$ curve of EIO cannot be described using a simple thermally activated model, as illustrated from the continuously varied gap $\Delta$ extracted if assuming the validity of the Arrhenius law $\rho(T) \propto exp(\Delta / k_B T)$ (inset of Fig.~\ref{Fig3}(a)). Similarly, $\rho(T)$ neither follows any of the variable-range hopping model $\rho \propto exp(T_0/T)^n$ ($n$ = 1/2, 1/3, or 1/4) \cite{Man2005}. 

Instead, in the range of 100-70 K, the resistivity follows a power-law relationship with temperature, namely, $\rho \propto T^{-\alpha}$ with the power-index $\alpha \approx$ 2.0 as displayed in Fig~\ref{Fig3}(b). Note, such a characteristic feature has been proposed theoretically and observed in another pyrochlore iridate Y$_2$Ir$_2$O$_7$ in the context of WSM, where the power-index $\alpha$ locates within the range 1 $\leq \alpha \leq$ 4, depending on the level of defects and the strength of correlations \cite{Hosur2012prl, sarma2015prb, rodionov2015prb, liu2020apl}. From this perspective, this observation may directly manifest the formation of a magnetic WSM state, consistent with what has been recently demonstrated in (111) EIO thin films \cite{Liu2021prl}. 

However, the $\rho(T)$ curve deviates from the power-law relationship at lower temperatures, indicating the inclusion of additional contributions to transport events. For topological semimetals with Dirac or Weyl nodes, the WAL effect is generally expected, because a peculiar Berry phase of $\pi$ is collected by carriers after circulating around the Fermi surface, inducing a destructive quantum interference between time-reversed loops formed by scattering trajectories \cite{zhou2016cpb}. In addition, theoretical studies based on the Feynman diagram analyses suggested that in a realistic interacting WSM system, the change of resistivity as a function of temperature is a combination of two competing terms \cite{lu2015prb}: 
\begin{equation}
\begin{aligned}
\Delta\sigma(T) = \sigma^{ee}(T) + \sigma^{WAL}(T) = C_{ee}\sqrt{T} - C_{WAL}T^{p/2}
\end{aligned}
\end{equation}
Here, $\sigma^{ee}(T)$ refers to the contribution from electron-electron interaction dressed by disorders; $\sigma^{WAL}(T)$ refers to the contribution from destructive quantum interference induced by the WAL effect (here $p$ = 3/2 for election-electron interaction \cite{lee1985rmp}). To verify this scenario, we fit the residual temperature-dependent conductivity $\Delta\sigma(T)$ achieved after subtracting the power-law contribution, as shown in Fig~\ref{Fig3}(c). The data are indeed well described using this model with the parameters $C_{ee}$ $\sim$0.08 and $C_{WAL}$ $\sim$0.03, signifying the essence of WAL as one of the characteristic signatures for WSMs \cite{kumar2023jap, zhang2020prb, huang2015prx, li2016np, lu2017fp}. 

\begin{figure}[htp]
\begin{center}
\includegraphics[width=0.65\textwidth]{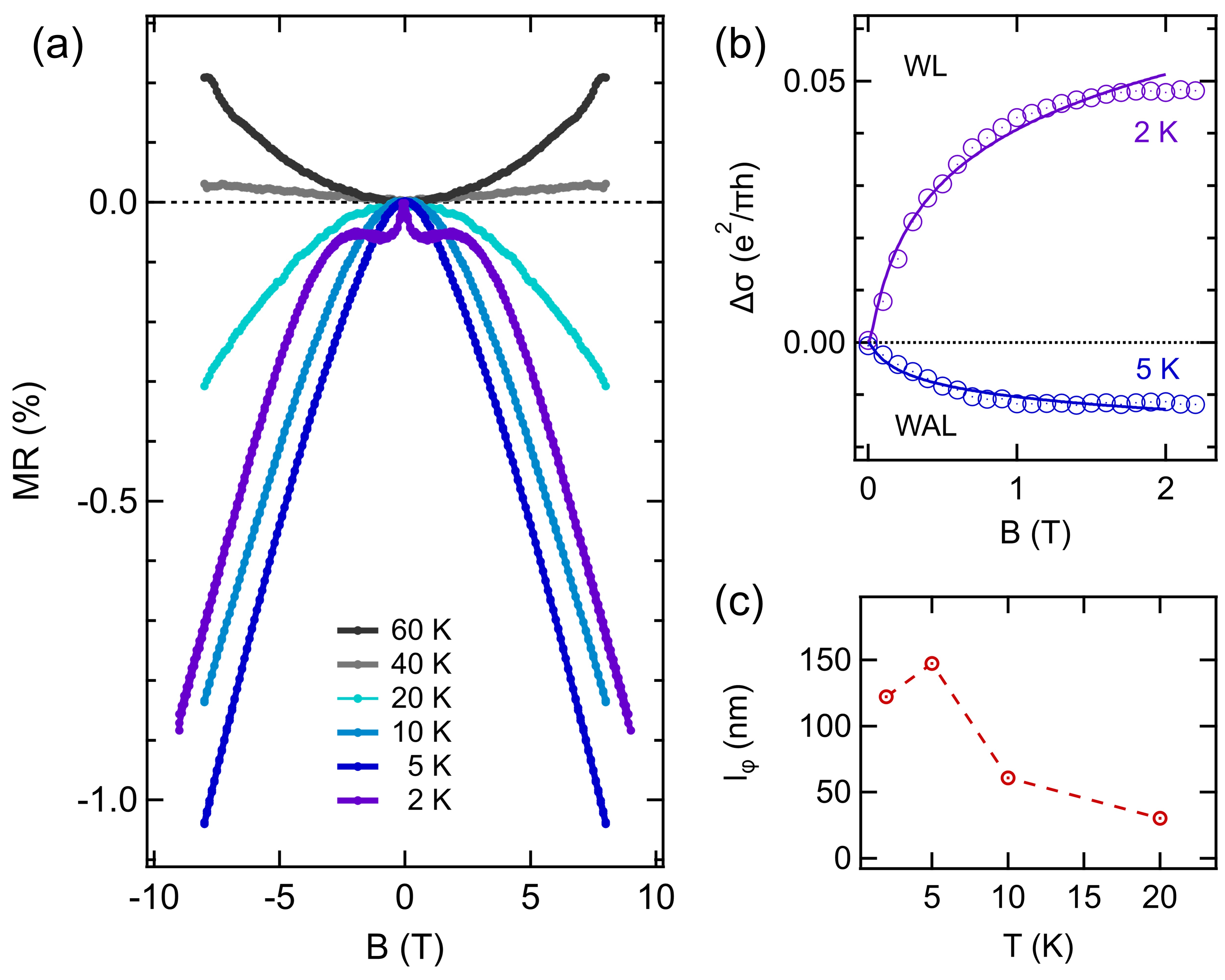}
\caption{\label{Fig4} (a) Data of magnetoresistance [MR = $\frac{{R(B)-R(0)}}{{R(0)}} \times 100\%$] recorded at a set of temperatures. The external magnetic field was applied along the out-of-plane [111] direction and the current along the in-plane [1$\bar{1}$0] direction of the film. (b) The representative magneto-conductance $\Delta\sigma$ (in units of $\frac{e^2}{\pi h}$) after polynomial subtraction at 2 K and 5 K, respectively. The solid curves are fits of the data using the HLN equation as described in the main text. (c) The characteristic length of phase coherence $l_{\varphi}$ extracted from the fits.}
\end{center}
\end{figure}

More insightful information regarding the WAL effect can be reached via field-dependent experiments. We measured the magnetoresistance (MR) of the (111) EIO films at low temperatures, with the magnetic field $B$ applied perpendicular to the sample surface. Note, each temperature spot was stabilized in zero field cooling, so as to reach a practically equal distribution of the two AFM domains. This can effectively minimize the asymmetric odd term in magnetoresistance (MR), given by one single domain with gigantic coercive fields \cite{fujita2015odd,Liu2021prl}. Then the longitudinal resistance was recorded while sweeping the magnetic field for a full cycle up to $\pm$8 T. Data extracted from symmetrization are shown in Fig.~\ref{Fig4}(a) for a few representative temperatures. At 60 K, the finite MR is positive and quadratic relative to $B$, primarily as a result of the cyclotron motion of carriers driven by the Lorentz force. The sign of MR is gradually switched from positive to negative at lower temperatures. This phenomenon is rather common and has also been reported in other (111) pyrochlore iridates thin films \cite{liu2020apl, fujita2015odd, fujita2018prm, kim2018prb}. Recap the non-coplanar ``all-in-all-out'' magnetic configuration on Ir pyrochlore sublattice, an external field along the [111] direction will cant the spins on the kagome planes slightly off the centers of tetrahedra for Zeeman energy. This can give rise to an overall field-induced net moment, suppressing the spin-flip scattering and generating a negative contribution to MR.          

Besides these classical terms that dominate the high-field behavior, the contribution induced by quantum interference may play more critical roles in the low-field regime, as evidenced by the cusp feature observed at 2 K near zero field. To further separate this contribution, we fit the corresponding magneto-conductance data within 3-8 T using a cubic function of $B$ to simulate the total contributions from field- and spin-related scattering:
\begin{equation}
\begin{aligned}
\sigma(B) = \Delta\sigma(B) + A_0 + A_1B + A_2B^2 + A_3B^3
\end{aligned}
\end{equation}
After subtracting the polynomials, we applied the Hikami-Larkin-Nagaoka (HLN) formula on the residual term $\Delta\sigma(B)$ to describe the interference-induced contribution under the strong spin-orbit approximation\cite{HLN1980}:     
\begin{equation}
\begin{aligned}
\Delta\sigma(B) = \alpha\frac{e^2}{\pi h}[\textrm{ln}(\frac{B_\varphi}{B}) - \psi(\frac{1}{2} + \frac{B_\varphi}{B})]
\end{aligned}
\end{equation}
where $\Psi$ is the digamma function; $\alpha$ is positive (negative) for WAL(WL); $B_\varphi$ = $\hbar/4el_{\varphi}^2$ is the characteristic field of phase coherence, and $l_{\varphi}$ is the corresponding phase coherence length, standing for the average distance that the electron can maintain its phase coherence. 

The data can be reasonably described using the HLN formula below 20 K (see Fig.~\ref{Fig4}(b) for two representative fits at 2 and 5 K, respectively). The extracted phase coherence length $l_{\varphi}$ displayed in Fig.~\ref{Fig4}(c) generally increases as temperature decreases due to the gradual suppression of inelastic scattering. In particular, $l_{\varphi}$ can reach about hundreds of nanometers below 5 K, much larger than the film thickness (and the mean free path) of around tens of nanometers, which is in accord with the requirement for quantum diffusion. However, as shown in Fig.~\ref{Fig4}(b), while the WAL effect is observed during 20-5 K, it is rather striking to reveal a crossover from WAL to WL below 5 K, and the WL effect is clearly established at 2 K. Notably, such a salient crossover has been exactly proposed theoretically \cite{lu2015prb}, which is reminiscent of the aforementioned competition between the interference-induced $\sigma^{WAL}$ and the interaction-induced $\sigma^{ee}$ that leads to a tendency to localization. 

In summary, we have achieved the fabrication of pyrochlore iridate EIO (111) epitaxial thin films via the reactive solid phase epitaxy method. Analyses on the temperature and the field dependences of resistivity reveal the presence of WAL at low temperatures, a characteristic signature of WSM that was ``buried'' by Ir magnetism. However, inclusion of the electron-electron interactions competes with the WAL effect, and triggers the crossover from WAL to WL below 5 K. These findings highlight the alternative route for pyrochlore iridate synthesis, and provide experimental insights about the interplay between quantum interference and many-body effects in correlated topological materials.\\

\begin{acknowledgments}
The authors acknowledge Y. Wang, X. Liu, S. Sobhit, and H. Kim for useful discussions. A portion of this work was carried out at the Synergetic Extreme Condition User Facility (SECUF). A portion of this work was based on the data obtained at beamline 1W1A of Beijing Synchrotron Radiation Facility (BSRF-1W1A). This work is supported by the National Key R\&D Program of China (Grant No. 2022YFA1403400, 2022YFA1403000, 2021YFA0718700), the National Natural Science Foundation of China (Grant No. 12204521, 12250710675, 11974409, 12104494), and the Strategic Priority Research Program of the Chinese Academy of Sciences (No. XDB33000000).\\ 
\end{acknowledgments}

\noindent{\textbf{DATA AVAILABILITY STATEMENT}}\\
The data that support the findings of this study are available from the corresponding author upon reasonable request.


\end{document}